\documentclass[aps,pre,twocolumn,superscriptaddress]{revtex4}
\usepackage{graphicx}
\usepackage{amsmath,amssymb}
\usepackage{float}
\usepackage{color}
\usepackage{bm}

\newcommand{\ignore}[1]{}
\newcommand{\beq}{\begin{equation}}
\newcommand{\eeq}{\end{equation}}
\newcommand{\mbold}[1]{\mbox{\boldmath $ #1 $}}

\begin{document}

\title
{Nontrivial temperature dependence of ferromagnetic resonance frequency for spin reorientation transitions}

\author{Masamichi Nishino}
\email[Corresponding author: ]{nishino.masamichi@nims.go.jp} 

\affiliation{Research Center for Advanced Measurement and Characterization, National Institute for Materials Science, Tsukuba, Ibaraki 305-0047, Japan}

\affiliation{Elements Strategy Initiative Center for Magnetic Materials, National Institute for Materials Science, Tsukuba, Ibaraki, Japan}

\author{Seiji Miyashita} 
\affiliation{Department of Physics, Graduate School of Science, The University of Tokyo, Bunkyo-Ku, Tokyo 113-0033, Japan}

\affiliation{Elements Strategy Initiative Center for Magnetic Materials, National Institute for Materials Science, Tsukuba, Ibaraki, Japan}

\begin{abstract} 
We find unusual temperature dependence of the ferromagnetic resonance (FMR) frequency $f_{\rm R}$ for the spin-reorientation (SR) transition, in which 
the easy axis changes depending on temperature, observed in the Nd permanent magnet, Nd$_2$Fe$_{14}$B: $f_{\rm R} \sim 0$ below the SR transition temperature ($T_{\rm R}$), drastic increase of $f_{\rm R}$ around $T_{\rm R}$, and decrease from a peak at higher temperatures. 
It is nontrivial that the SR transition causes the unusual behavior of the FMR frequency in a wide temperature region. 
We show the mechanism of the temperature dependence by theoretical and computational analyses. 
We derive a general relation between $f_{\rm R}$ and magnetizations to help the understanding of the mechanism, and clarify that the fluctuation of the transverse magnetization is a key ingredient for the resonance in all temperature regions. 
\end{abstract}

\maketitle

Ferromagnetic resonance (FMR) measurement is one of the most important methods to study the dynamics of ferromagnetic materials. FMR frequency signals tell us the information about spin dynamics such as spin precession motion, damping factor, etc. However, how the FMR frequency behaves for spin-reorientation (SR) transitions is hardly understood. 
The neodymium (Nd) permanent magnet~\cite{Sagawa,Herbst2,Coey}, Nd$_2$Fe$_{14}$B, which has strong coercivity, is an indispensable material in modern technologies, applied to information-storage devices, hybrid and electric vehicles, generators, etc. This magnet exhibits a temperature-induced spin-reorientation (SR) transition. 
However, due to its strong magnetic anisotropy energy~\cite{Kato}, it is difficult to observe FMR signals in such a hard magnet, and the spin dynamics on the SR transition in the magnet has not been clarified. The temperature dependence of the FMR frequency is an open question. 

Time domain measurement (TDM) of spin dynamics has rapidly been developed for the coherent control of spin precession motion toward spintronics technologies. 
The change of the precession frequency accompanying a SR transition has been detected recently in a rare-earth orthoferrite (RFeO$_3$) by using time-resolved magneto-optical effect microscope~\cite{Kimel,Yamaguchi1,Yamaguchi2}. 
Very recently a TDM of spin dynamics has been performed to evaluate the interlayer exchange coupling between hard and soft Nd magnet layers, and the resonance frequency for a single thick layer of the hard Nd magnet was estimated to be $f=\omega/(2\pi)=161$ GHz under an external field at 2 T at room temperature~\cite{Mandal}. The SR transition in the Nd magnet is a new target for TDM. 
Precession motion associated with SR transitions becomes an important and attractive topic. 

Motivated by the experimental situation, we study the dynamics of 
the SR transition for the Nd magnet in the present Rapid Communication. 
We find that the FMR frequency exhibits nontrivial temperature dependence with a drastic change around $T_{\rm R}$. 
We investigate the origin of the behavior and find a universal mechanism for systems with the SR transition.   

In the present work we study an atomistic spin model for the Nd magnet~\cite{Toga,Nishino,Hinokihara}. 
Compared to micromagnetics modeling, i.e., coarse-grained continuum model approach often used for studies of permanent magnets~\cite{Kronmullar}, the atomistic spin modeling has advantages for investigating microscopic magnetic properties and thermal fluctuation effects~\cite{Hinzke2,Ostler,Evans,Nishino2}.  
Indeed temperature dependences of magnetization, domain wall (DW) width, DW free energy, anisotropy effects of Nd and Fe, etc. have been clarified by using the atomistic spin model for the Nd magnet (see \eqref{Nd-mag-model})~\cite{Toga,Nishino,Hinokihara,Miyashita,Toga2}, in which the microscopic parameters were taken mainly from first-principles calculations. 
Here we focus on the temperature dependence of the FMR frequency at zero external field. Under a finite external field $h_{\rm e}$, the resonance frequency shifts by $\gamma h_{\rm e}/(2\pi)$.

The atomistic Hamiltonian for the Nd magnet is given by the form~\cite{Toga,Nishino,Hinokihara}: 
\beq
{\cal H}= - \sum_{i < j} 2 J_{ij} \mbold{s}_i \cdot \mbold{s}_j - \sum_i^{\rm Fe} D_i (s_i^z)^2 + \sum_{i}^{\rm Nd} \sum_{l,m} \Theta_{l,i} A_{l,i}^m \langle r^l\rangle_i \hat{O}_{l,i}^m. 
\label{Nd-mag-model}
\eeq
Here $J_{ij}$ denotes the Heisenberg exchange coupling between the $i$th and $j$th sites, and $D_i$ is the magnetic anisotropy constant for Fe atoms. 
For Fe and B atoms, $\mbold{s}_i$ is the magnetic moment at the $i$th site. For Nd atoms, $\mbold{s}_i$ is the moment of the valence (5d and 6s) electrons and is coupled antiparallel to the moment of the 4-f electrons $\mbold{\cal J}_i$, and thus the total magnetic moment is given by $\mbold{S}_i=\mbold{s}_i + \mbold{\cal J}_i$, in which $\mbold{\cal J}_i=g_{\rm T}\mbold{J}_i \mu_{\rm B}$ with the magnitude of the total angular momentum, $J=9/2$, and Land\'e g-factor, $g_{\rm T}=8/11$. For Fe and B atoms, $\mbold{S}_i= \mbold{s}_i$ is defined. The Zeeman term for the model is given as $-\sum_i \mbold{h}_{\rm e} \cdot \mbold{S}_i$. 

The last term of Eq.~\eqref{Nd-mag-model} is the magnetic anisotropy energy for Nd atoms, where $\Theta_{l,i}$,  $A_{l,i}^m$, $\langle r^l\rangle_i$, and $\hat{O}_{l,i}^m$ are the Stevens factor, the coefficient of the spherical harmonics of the crystalline electric field, the average of $r^l$ over the radial wave function, and Stevens operator, respectively, at site $i$ for Nd atoms. 
The summation for $l$ runs $l=2,4,6$ and diagonal operators ($m=0$) give dominant contribution, and then the last term is given in the series of $J_i^z$ as 
\beq
 \sum_{i}^{\rm Nd} D_1^{\rm Nd} (J_i^z)^2+D_2^{\rm Nd} (J_i^z)^4+D_3^{\rm Nd} (J_i^z)^6 + const.
\label{Aniso_ene}
\eeq

\begin{figure}[t!]
  \begin{center}
     \includegraphics[width=9cm]{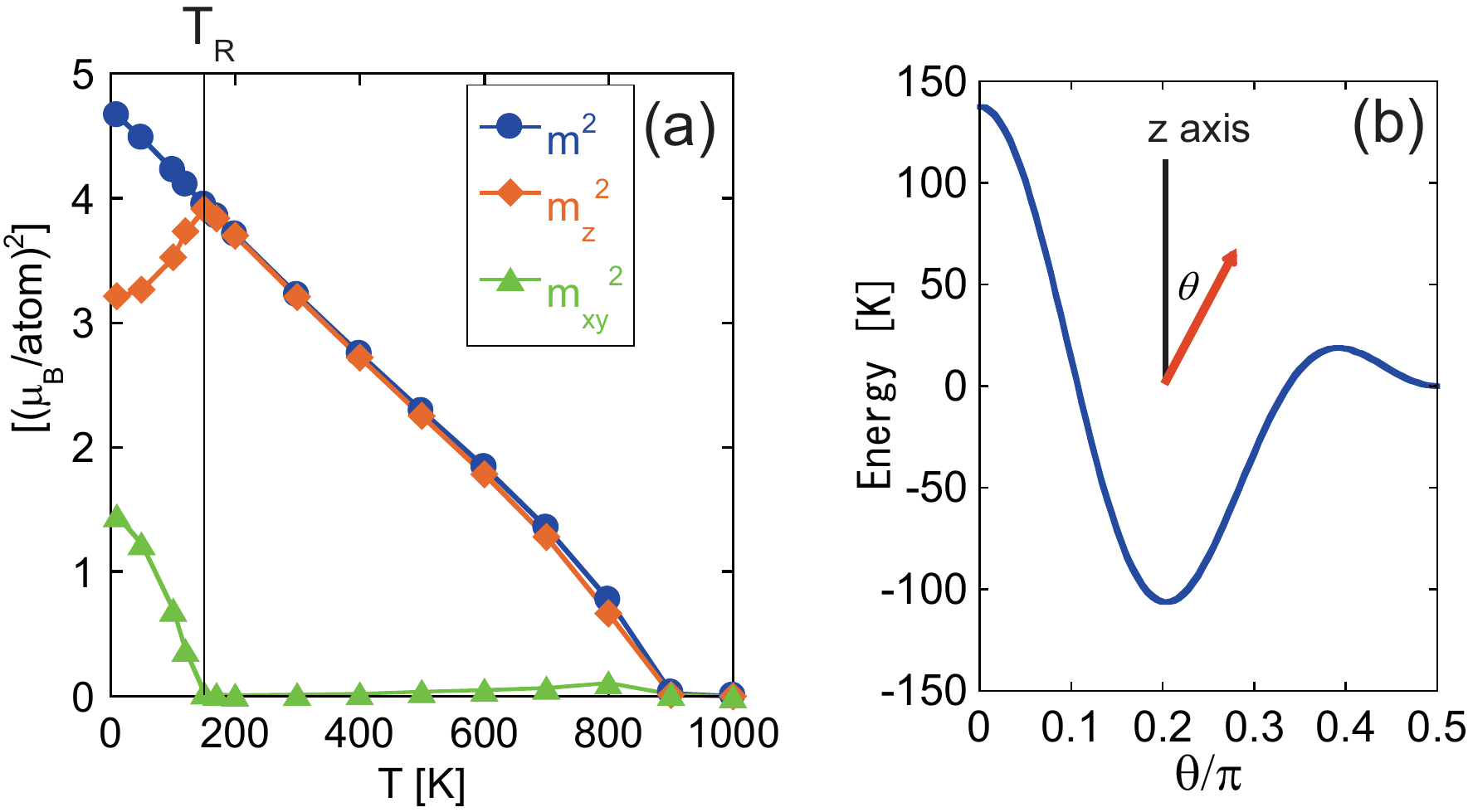}
  \end{center}
\caption{(color online) (a) Temperature dependences of $m^2$, $m_z^2$, and $m_{xy}^2$ for the Nd magnet model \eqref{Nd-mag-model}. (b) $\theta$ dependence of 
the per-site anisotropy energy of Eq.~\eqref{Aniso_ene} for the Nd magnet model.}
\label{Fig_Nd_mag_pot}
\end{figure}

Figure~\ref{Fig_Nd_mag_pot} (a) depicts the temperature dependences of $m^2 \equiv  \frac{1}{N^2} \langle \sum_{\alpha=x,y,z}( \sum_i^N  S_i^\alpha )^2 \rangle$, $m_z^2 \equiv  \frac{1}{N^2}  \langle (\sum_i^N  S_i^z )^2 \rangle$, and $m_{xy}^2 \equiv  \frac{1}{N^2} \langle (\sum_i^N  S_i^x )^2+ ( \sum_i^N   S_i^y)^2  \rangle$, where $\langle X \rangle$ is the thermal average of $X$ for the system with $6\times6\times6$ unit cells with periodic boundary conditions~\cite{MC}. The SR transition occurs at $T_{\rm R} \sim 150$ K~\cite{Toga,Nishino}, close to the experimental estimation~\cite{Hirosawa2,Andreev,Kou}. The critical temperature is $T_{\rm c} \sim 850$ K, a little bit larger than the experimental estimation~\cite{Hirosawa2,Andreev,Kou}, which does not affect the discussion here. 
The spins align tilted from the c axis ($z$ axis) for $T<T_{\rm R}$, while they align parallel to the c axis for $T>T_{\rm R}$. 
Because the minimum of the anisotropy energy \eqref{Aniso_ene} is realized at $\theta \simeq 0.2\pi$ (Fig.~\ref{Fig_Nd_mag_pot} (b), see also Fig.~S1 in Supplemental Material (SM)~\cite{Suppl}), the most preferable direction of the magnetization in the ground state is not the c-axis ($z$-direction), which is the easy axis at room temperature.

Now we study the temperature dependence of the FMR frequency at zero external field. The FMR spectrum $I(f)$ is calculated by the auto-correlation function of spins (power spectrum):
\beq
I^{\alpha} (f) \equiv 
\dfrac{1}{T}\int^{t_{0}+T}_{t_{0}} d\tau I^\alpha (\tau) {\rm e}^{i 2\pi f \tau},
\eeq 
where 
\beq
I^{\alpha } (\tau) =\dfrac{1}{T}\int^{t_{0}+T}_{t_{0}}dt \langle \Bar{S}^{\alpha } \left( t\right) \Bar{S}^{\alpha } \left(t+ \tau \right) \rangle. 
\eeq
Here $\Bar{S}^{\alpha }=\dfrac {1}{N}\sum _{i}S_{i}^{\alpha }$ for $\alpha=x$, $y$, and $z$.  
For the spin dynamics, we adopt the stochastic Landau-Lifshitz-Gilbert (SLLG) equation~\cite{Nishino,Garcia}: 
\begin{align}
 \frac{d}{d t} \bm{S}_i &= -\frac{\gamma}{1+\alpha_i^2}  \bm{S}_i \times   (\bm{H}_i^{\rm eff} + \bm{\xi}_i  )   \nonumber  \\ 
&- \frac{\alpha_i \gamma}{(1+\alpha_i^2)S_i}  \bm{S}_i \times  (\bm{S}_i \times (\bm{H}_i^{\rm eff}+ \bm{\xi}_i) ), 
\label{LLG-noise}
\end{align} 
where $\alpha_i$ is the damping factor at the $i$th site and $\gamma$ is the gyromagnetic constant. Here $\bm{H}_i^{\rm eff}=-\frac{\partial \cal H}{\partial \bm{S}_i}$ is the effective field applied at the $i$th site from the exchange interactions and the anisotropy terms, and $\bm{\xi}_i $ is a random noise introduced into each site and has the relation:
$\langle \xi_k^\mu(t) \rangle=0$,  $\langle \xi_k^\mu(t)\xi_l^\nu (s) \rangle=2\tilde{D}_k \delta_{kl}\delta_{\mu \nu} \delta(t-s)$. 
Here $\xi_i^\mu$ is the $\mu$(=$x$,$y$ or $z$) component of the white Gaussian noise. We assume $\alpha_i=0.1$~\cite{Kronmullar}. 
The strength of the noise is given by the fluctuation-dissipation relation: $\tilde{D}_i=\frac{\alpha}{S_i} \frac{k_{\rm B}T}{\gamma}$, which guarantees the thermal equilibrium state in the steady state. 

\begin{figure}[t!]
  \begin{center}
     \includegraphics[width=6cm]{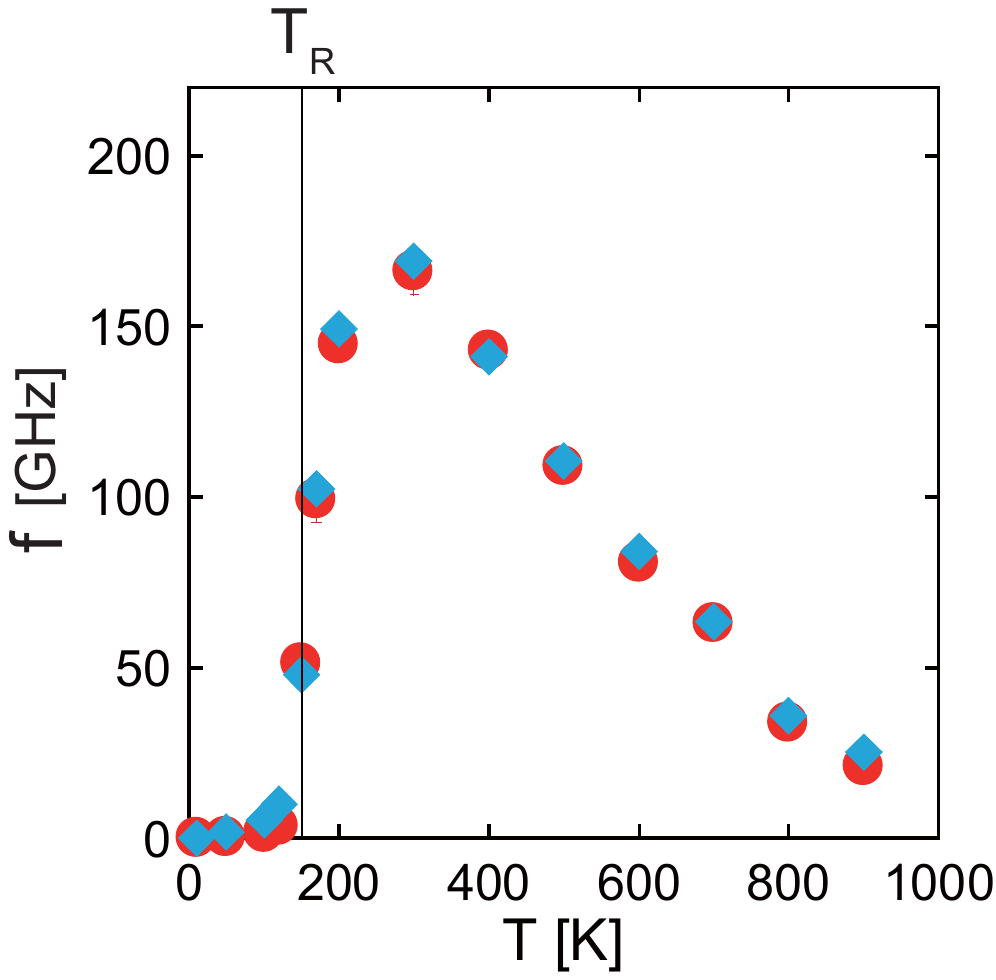}
  \end{center}
\caption{(color online) Temperature dependences of $f_{\rm R}$ (blue diamonds) and $f_{\rm SW}$ (red circles) for the Nd magnet model. }
\label{Fig_f_Nd}
\end{figure}

The temperature dependence of the FMR frequency $f_{\rm R}$ is plotted by blue diamonds 
in Fig.~\ref{Fig_f_Nd}. The red circles are explained below. 
We find that $f_{\rm R}$ exhibits a drastic change around $T_{\rm R}$ and non monotonic temperature dependence. 
In particular, it is found that $f_{\rm R} \sim 0$ for the magnetization tilted from the c axis and $f_{\rm R}$ drops sharply at the SR point. The spectrum of $\bar{I}(f) \equiv \frac{I^{x}(f)+I^{y}(f)}{2}$ has a single peak (Fig.~S2 in SM~\cite{Suppl}), and $f_{\rm R}$ was estimated by the median of the spectrum of 
$\bar{I}(f)$~\cite{calc_for_I}. The estimated $f_{\rm R}$ at 400 K ($T/T_{\rm c} \sim 0.5$) in the present simulation is 141 GHz under zero field and then $f_{\rm R}=197$ GHz under 2T field, which is the same order of the experimental value (161GHz) at room temperature~\cite{Mandal}.

The sudden decrease of the resonance frequency is related to the SR transition. To capture the fundamental mechanism of the non-monotonic temperature dependence of $f_{\rm R}$, the Nd magnet model~\eqref{Nd-mag-model} is too complicated, and thus we introduce a minimal model which exhibits the SR transition: 
\beq
{\cal H}= - J \sum_{\langle i ,j \rangle} \mbold{S}_i \cdot \mbold{S}_j -D_1 \sum _i (S_i^z)^2 - D_2 \sum _i (S_i^z)^4, 
\label{Ham_D1D2}
\eeq
where $J=1$ is the exchange constant for nearest-neighbor pairs, used as the unit of energy, $D_1$ and $D_2$ are primary and secondary uniaxial anisotropy constants, respectively, and $\mbold{S}_i$ is a classical unit spin. 

In the ground state (GS) all magnetic moments are aligned in the same direction and the per-site exchange energy is $E_{\rm ex}=-\frac{z}{2}J$, where $z$ is the coordination number. Thus the per-site GS energy is given by the minimum of 
$E_{\rm G}=-\frac{z}{2}J-D_1(S^z)^2-D_2(S^z)^4$. 
For $D_2=0$ the easy axis is given by $\theta_0=0$. 
However, for $D_2 < 0$, the magnetic moments are aligned in the direction of 
$\theta_0=\cos^{-1} \Big(\sqrt{-\frac{D_1}{2D_2}} \Big)$.
Here we adopt $D_1=1$ and $D_2=-0.7$ to have $\theta_0\simeq 0.2\pi$, 
which shows a similar potential minimum as in the Nd magnet model \eqref{Nd-mag-model} (see Fig.~S3 in SM~\cite{Suppl}).

\begin{figure}[t!]
  \begin{center}
     \includegraphics[width=9cm]{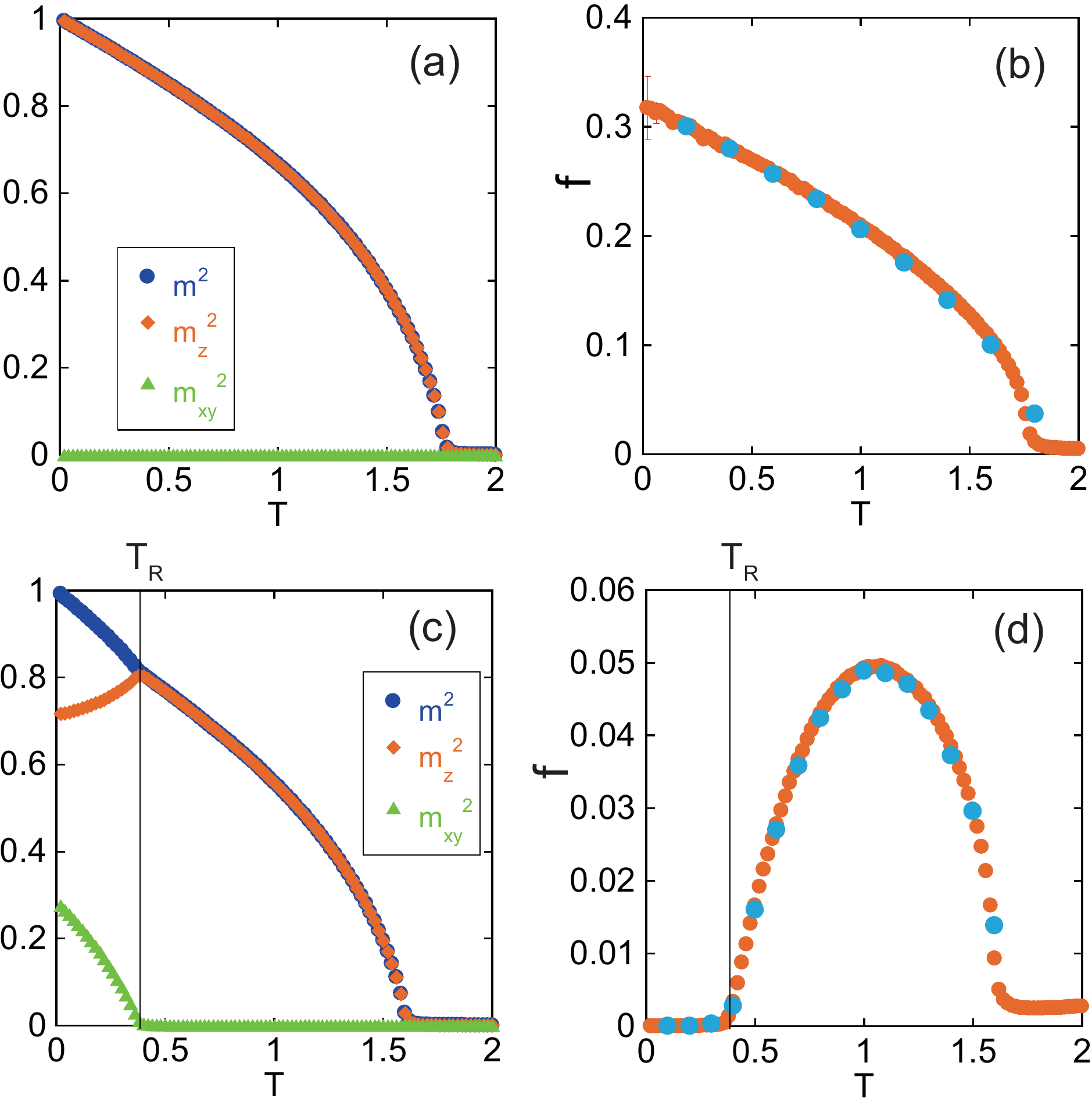}
  \end{center}
\caption{(color online) 
For $D_1=1$ and $D_2=0$ in the minimal model \eqref{Ham_D1D2}, temperature dependences of $m^2$, $m_z^2$, and $m_{ xy}^2$ are given in (a) and those of $f_{\rm R}$ (blue circles) and $f_{\rm SW}$ (red circles) are plotted in (b). 
For $D_1=1$ and $D_2=-0.7$, the former and latter dependences are depicted in (c) and (d), respectively. $k_{\rm B}=1$ was set. The time step for Eq.~\eqref{LLG-noise} $dt=0.005$ was used for the evaluation of $f_{\rm R}$.}
\label{Fig_D1D2_models}
\end{figure}

As a reference for the conventional case, we first check the temperature dependence of $f_{\rm R}$ for the simple ferromagnet with $D_1=1$ and $D_2=0$. 
The temperature dependences of $m^2$, $m_z^2$, and $m_{xy}^2$ ($k_{\rm B}=1$ is set) are depicted in Fig.~\ref{Fig_D1D2_models} (a), where $T_{\rm C}\simeq 1.77J$, which is a little bit higher than that for the Heisenberg model ($1.44 J$)~\cite{Peczak}. The resonance frequency $f_{\rm R}(T)$ is plotted by blue circles in Fig.~\ref{Fig_D1D2_models} (b) (see also Fig.~S4 in SM~\cite{Suppl}), which shows a monotonic temperature dependence similar to that of $m_z^2$.

Next, we study the case with the SR transition with $D_2=-0.7$.  
We find a SR transition at $T_{\rm R}\simeq 0.4$ and ferromagnetic transition at $T_{\rm C}\simeq 1.6$ in Fig.~\ref{Fig_D1D2_models} (c). 
In Fig.~\ref{Fig_D1D2_models} (d) $f_{\rm R}(T)$ is shown (see also Fig.~S5 in SM~\cite{Suppl}). 
We find $f_{\rm R} \sim 0$ below $T_{\rm R}$, $f_{\rm R}$ increases above $T_{\rm R}$ with rising temperature, and it exhibits a peak at an intermediate temperature and reduction at higher temperatures. 
This behavior is qualitatively similar to that found in the Nd magnet model. 
Thus, $f_{\rm R} \sim 0$ at low temperatures is attributed to the SR transition. 

Now we analyze the mechanism of $f_{\rm R} \sim 0$ in the low temperature phase. 
Below $T_{\rm R}$ the magnetization is tilted from the c-axis, 
where 
the effective field applied to each site is given as 
\beq
\mbold{h}_{i, {\rm eff}}= -\frac{\partial{H}}{\partial \mbold{S}_i}= J \sum_{j}^{\rm NN} \mbold{S}_j + h_{\rm eff}^{\rm Aniso} \mbold{e}_z, 
\eeq
where 
$h_{\rm eff}^{\rm Aniso} = 2 D_1 S_i^z+4D_2 (S_i^z)^3$
is the contribution from the anisotropy term. 
The field from the exchange interactions does not contribute to the precession motion because it is parallel to the spin alignment, 
and thus the frequency for the precession motion is given by 
\beq
f=\gamma h_{\rm eff}^{\rm Aniso}/(2\pi).
\eeq

For $D_2=0$, $f=\gamma h_{\rm eff}^{\rm Aniso}/(2\pi) =2 D_1 m_z/(2\pi) =0.318$ (see Fig.~\ref{Fig_D1D2_models} (b)), which depends on the temperature proportionally to $m_z$. It is the conventional temperature dependence. 
On the other hand, for $D_2 \ne 0$, the situation is different. 
Considering the relations 
$\frac{dE_{\rm G}}{d\theta}|_{\theta=\theta_0}=\frac{dE_{\rm G}}{dS_z} \frac{dS_z}{d\theta}|_{\theta=\theta_0}=- h_{\rm eff}^{\rm Aniso} \sin \theta_0$,
 we notice a relation: $h_{\rm eff}^{\rm Aniso}=0$ for $\theta=\theta_0$ because $\frac{dE_{\rm G}}{d\theta}|_{\theta=\theta_0}=0$ and $\sin \theta_0\ne 0$. 
Thus, we have an important consequence:  
\beq
f=\gamma h_{\rm eff}^{\rm Aniso}/(2\pi)=0 \quad {\rm for}\quad \theta=\theta_0\ne 0.
\label{f_zero}
\eeq

To deduce Eq.~\eqref{f_zero}, non-zero $\theta_0$ is essential. Here we adopted $-D_1 \sum_i (S_i^z)^2$ and $-D_2 \sum_i (S_i^z)^4$ anisotropy terms for the minimal model. Instead of this choice, for example, we can adopt $-D_2 \sum_i (S_i^z)^4$ and $-D_3 \sum_i (S_i^z)^6$, in which $D_2>0$ and $D_3<0$, without the $D_1$ term. This choice also can give non-zero $\theta_0$.

Finally we investigate the temperature dependence of $f_{\rm R}$ for $T>T_{\rm R}$. For $T>T_{\rm R}$, $m_{xy}^2$ does not appear and the $D_2$ term is not essential. Thus we consider the situation for $D_2=0$ with applying Stoner-Wohlfarth's single domain picture. 
The free energy for the uniform single domain with a magnetic anisotropy $K$ under zero field is given as $F=-K(T) \cos^2 \theta$, 
where $K$ is a function of temperature. 
Here $\cos \theta=\frac{M_z}{M}$. $M$ and $M_z$ are the total magnetization and its $z$ component, respectively. 
It should be noted that this $\theta$ is not the same as $\theta_i$ defined for each spin as $S_i^z=S \cos \theta_i$. 
The internal field for each magnetic moment is  
\beq
h_{\rm eff}=-\frac{dF}{dM_z}=2K(T)\frac{M_z}{M^2}.
\eeq
Thus the resonance frequency is given as $f_{\rm SW}=\gamma h_{\rm eff}/(2 \pi)=\gamma K \frac{M_z}{M^2}/\pi$.

To evaluate $K(T)$ in the microscopic models \eqref{Nd-mag-model} and \eqref{Ham_D1D2},
we derive the following relation between $K(T)$ and the transverse-field susceptibility $\chi_{xx}$.
The free energy of the system under a transverse-field is  
$F=-K \cos^2 \theta -H_x M \sin \theta$.
Here 
$M_x=M \sin \theta_{\rm min}=\frac{M^2}{2K} H_x$, 
where $\theta_{\rm min}$ is the angle to realize the stable state. 
$M_x$ is also expressed as $M_x=\chi_{xx} H_x$, where $\chi_{xx}$ is the susceptibility in the hard direction.
From these, we have $K(T)=\frac{M^2}{2\chi_{xx}}$. 
The susceptibility $\chi_{xx}$ is evaluated by 
$\chi_{xx}=\frac{N^2  m_{xy}^2 }{2 k_{\rm B} T}$. 
Then we find a relation: $K(T)=\frac{k_{\rm B} T M^2 }{N^2  m_{xy}^2}$, from which
we obtain the temperature dependence of the resonance frequency as 
\beq
f_{\rm SW}=\frac{1}{\pi}\gamma K \frac{M_z}{M^2}=\frac{1}{\pi}\gamma \frac{k_{\rm B} T  m_z}{N m_{xy}^2}. 
\label{f_sw}
\eeq

We notice that this formula can be extended for lower temperatures below $T_{\rm R}$.  Because of $m_z \sim O(1)$ and $m_{xy}^2 \sim O(1)$ for $T<T_{\rm R}$,  
$f_{\rm SW}=\gamma k_{\rm B} T \times O(1/N) \simeq 0$. It is also worth noting that 
 for $T>T_{\rm R}$, $m_{xy}^2 \sim O(1/N)$ and $f_{\rm SW} \sim O(1)$. 
In Fig.~\ref{Fig_f_Nd}, Fig.~\ref{Fig_D1D2_models} (b), and Fig.~\ref{Fig_D1D2_models} (d),  $f_{\rm SW}$ are plotted by red circles. 
We find excellent agreements between $f_{\rm SW}$ and $f_{\rm R}$ in all cases including the temperature region $T<T_{\rm R}$.

\begin{figure}[t!]
  \begin{center}
     \includegraphics[width=9cm]{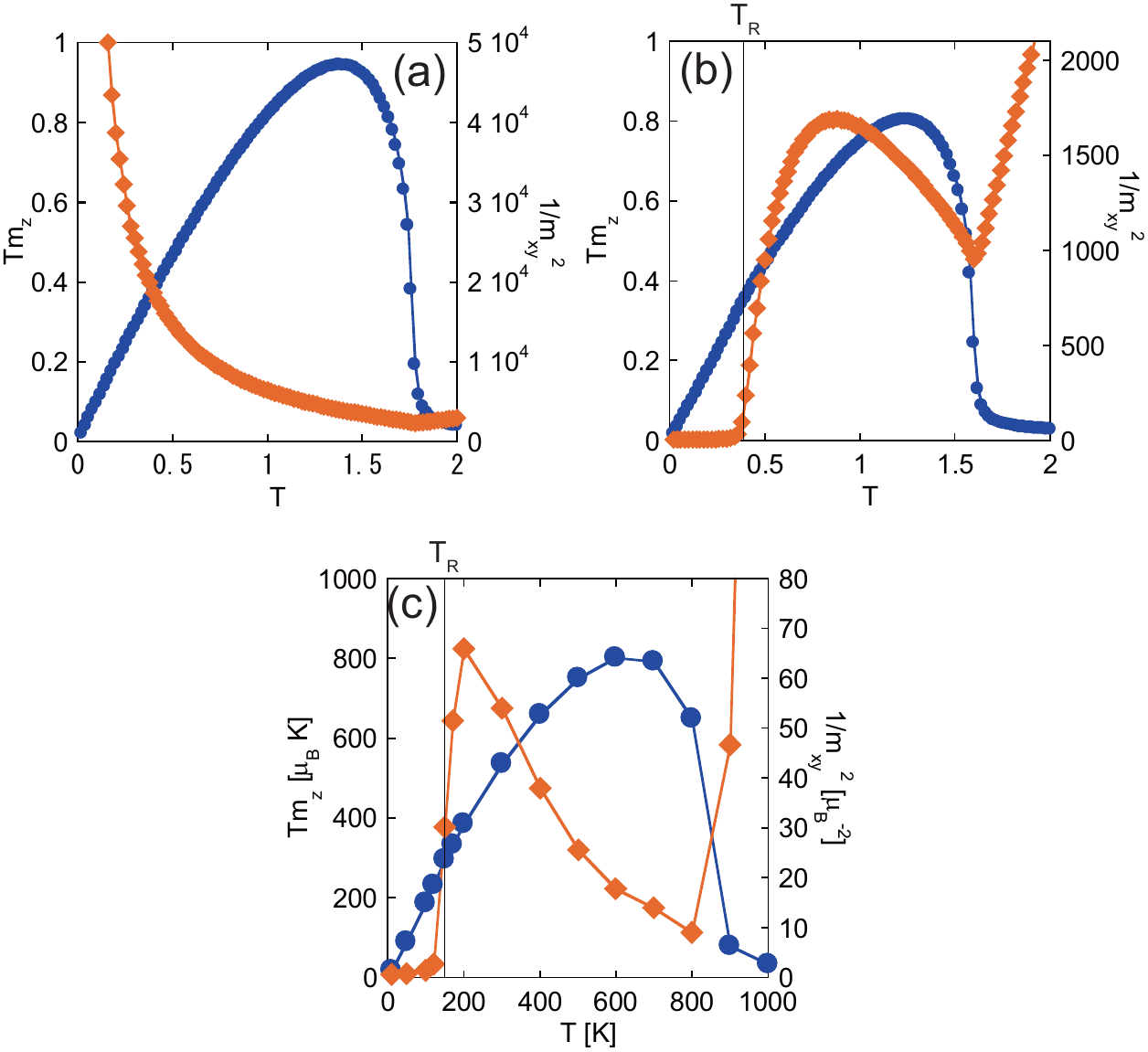}
  \end{center}
\caption{(color online) Temperature dependences of $T m_z$ (blue circles) and $1/m_{xy}^2$ (red diamonds) for the minimal model \eqref{Ham_D1D2} with (a) $D_2=0$ and (b) $D_2=-0.7$, and for (c) the Nd magnet model \eqref{Nd-mag-model}. }
\label{Fig_Tmz_vs_mxy2-inv}
\end{figure}

Figures~\ref{Fig_Tmz_vs_mxy2-inv} (a), (b), and (c) depict the temperature dependences of $T m^z$ and $1/m_{xy}^2$ for model \eqref{Ham_D1D2} with $D_2=0$, $D_2 \ne 0$, and model \eqref{Nd-mag-model}, respectively. 
We find that the temperature dependence of $T m^z$ is qualitatively similar in all cases, while that of $1/m_{xy}^2$ is qualitatively different, and the behavior of $m_{xy}^2$, i.e., the fluctuation of the transverse magnetization, is a very important ingredient for the FMR frequency.  

To conclude, we showed unusual temperature dependence of the FMR frequency in a wide temperature region for the spin-reorientation transition in the Nd magnet, caused by the competition between magnetic anisotropy energies: $f_{\rm R} \sim 0$ below $T_{\rm R}$,  $f_{\rm R}$ drastically changes around $T_{\rm R}$, and it exhibits a peak and then decrement at higher temperatures. This is totally different from the dependence of conventional magnets with a single uniaxial anisotropy energy, in which a monotonic decrease of the FMR frequency is observed. 
We clarified the mechanism for $f_{\rm R} \sim 0$ below $T_{\rm R}$.  
It is worth noting that the state of the tilted spin alignment is stable, and the precession around the GS easy axis (parallel to the GS magnetic moments) does not occur. 
We also derived the formula \eqref{f_sw} for the FMR frequency above $T_{\rm R}$ in connection to the temperature $T$, magnetization along the easy axis $m_z$, and the fluctuation of the magnetization along the hard axis (hard plane) $m_{xy}^2$. 
We found that this formula is  a good description for overall temperature region, and $m_{xy}^2$ in the formula is important for the qualitative nature of the FMR frequency. This formula is generally applied to other materials with SR transitions. 

This finding of the unusual temperature dependence of the FMR frequency will stimulate time-domain measurement studies on precession motions in SR transitions and spin-state switching such as ultrafast optical control of precession motion for RFeO$_3$~\cite{Kimel,Yamaguchi2} and of spin state for epsilon-Fe$_2$O$_3$~\cite{ohkoshi-JACS} and microwave-assisted magnetic recording for Co/Pd layer, etc~\cite{Recording,Nozaki}. 

We would like to thank Dr. Hirosawa, Dr. Mandal, and Dr. Takahashi for instruction of experimental features of the Nd magnet and helpful discussions. 
The present work was supported by the Elements Strategy Initiative Center for Magnetic Materials under the outsourcing project of MEXT.


\begin{thebibliography}{99}
\bibitem{MC}
We used a Monte Carlo (MC) method for the estimation of $m^2$, $m_z^2$, $m_{xy}^2$ and $m_z (\equiv \frac{1}{N} \langle |\sum_i^N  S_i^z | \rangle$) in Eq.~\eqref{f_sw}, which were obtained from five independent simulations of 200,000 MC steps for equilibration and another 1,000,000 MC steps for measurement. 
The SLLG and MC methods lead to the same equilibrium state and the latter can obtain static quantities faster than the former~\cite{Nishino}.
\bibitem{calc_for_I}
$\bar{I}(f)$ was evaluated from 16 independent simulations. 
In each simulation, the time step for Eq.~\eqref{LLG-noise} $dt=0.1$ fs was set and 1,000,000 steps for equilibration and another 5,000,000 steps for measurement were used.
\end{thebibliography}
\end{document}